\documentclass[letterpaper,10 pt,conference]{ieeeconf}

\newtheorem{theorem}{Theorem}
\newtheorem{definition}{Definition}

\usepackage{amssymb,amsmath,dsfont}
\usepackage{mathrsfs}
\usepackage{lipsum}
\usepackage{mathtools}

\usepackage[latin1]{inputenc}
\usepackage{graphicx}
\usepackage{psfrag}
\usepackage{color}
\usepackage{cite}
\usepackage{enumerate}
\usepackage{subcaption}
\usepackage{algorithm}
\usepackage[noend]{algpseudocode}
\usepackage{hyperref}

\makeatletter
\renewcommand*\env@matrix[1][*\c@MaxMatrixCols c]{%
  \hskip -\arraycolsep
  \let\@ifnextchar\new@ifnextchar
  \array{#1}}
\makeatother

\DeclareMathOperator*{\argmin}{arg\,min}

\IEEEoverridecommandlockouts                              

\title{\LARGE \bf
Data-Driven Scenario Optimization for Automated Controller Tuning with Probabilistic Performance Guarantees
}

\author{Joel A. Paulson and Ali Mesbah
\thanks{J. A. Paulson is with the Department of Chemical and Biomolecular Engineering, The Ohio State University, Columbus, OH 43210, USA. {\tt\small paulson.82@osu.edu}}%
\thanks{A. Mesbah is with the Department of Chemical and Biomolecular Engineering, University of California, Berkeley, CA 94720, USA. {\tt\small mesbah@berkeley.edu}}%
}

\begin{document}

\maketitle
\thispagestyle{empty}
\pagestyle{empty}

\begin{abstract}
Systematic design and verification of advanced control strategies for complex systems under uncertainty largely remains an open problem. Despite the promise of black-box optimization methods for automated controller tuning, they generally lack formal guarantees on the solution quality, which is especially important in the control of safety-critical systems. This paper focuses on obtaining closed-loop performance guarantees for automated controller tuning, which can be formulated as a black-box optimization problem under uncertainty. 
We use recent advances in non-convex scenario theory to provide a distribution-free bound on the probability of the closed-loop performance measures. To mitigate the computational complexity of the data-driven scenario optimization method, we restrict ourselves to a discrete set of candidate tuning parameters.
We propose to generate these candidates using constrained Bayesian optimization run multiple times from different random seed points. We apply the proposed method for tuning an economic nonlinear model predictive controller for a semibatch reactor modeled by seven highly nonlinear differential equations.
\end{abstract}

{\keywords Automated controller tuning, nonconvex scenario optimization, constrained Bayesian optimization.}

\section{Introduction}
\label{sec:intro}

Significant progress has been made in both the formulation and design of optimization- and learning-based controllers that can deal with multivariate dynamics, constraints, and uncertainties in the system and environment. 
However, systematic design of such advanced model-based controllers remains an open challenge since several tuning parameters must be selected by users.
These tuning parameters may consist of continuous, discrete, and/or categorical variables, and often enter the control design problem in non-smooth and non-convex ways. Therefore, in practice, they are generally chosen through extensive trial-and-error simulation or experimentation \cite{blondel2000survey}. This approach, however, is challenging for uncertain nonlinear systems as no closed-form solution exists for even controller \textit{verification} \cite{blondel2000survey,paulson2018shaping}, much less design. Thus, there has been a growing interest in new paradigms that can \textit{automatically} tune complex control structures for nonlinear systems under uncertainty \cite{berkenkamp2016safe,piga2019performance,fiducioso2019safe,lu2020mpc,konig2020safety}.

Data-driven optimization methods have become powerful tools for automated design in several application areas \cite{shahriari2015taking}. Bayesian optimization (BO) is one of the most successful approaches for \textit{black-box} optimization due to its data efficiency. As such, BO is useful whenever the objective function is expensive to evaluate, one does not have access to derivative information, and when the objective is non-convex with many local optima \cite{shahriari2015taking,snoek2012practical,marchant2012bayesian}. Recently, BO has been applied to tuning of model predictive control (MPC) \cite{piga2019performance,lu2020mpc} and other control architectures \cite{fiducioso2019safe,konig2020safety}. Although promising results observed in practice, these works lack formal guarantees on the solution quality. Such guarantees are especially important for safety-critical systems in which the closed-loop system must satisfy constraints and performance requirements despite uncertainty. 

The key contribution of this paper is to present an automated controller tuning method that provides guarantees on closed-loop performance and constraint satisfaction. We first pose the tuning problem as a black-box optimization with robust constraints, which can be tackled with constrained variants of BO \cite{hernandez2016general}. However, an important complication is that the expected performance cost and worst-case constraint violation cannot be computed exactly -- these must be estimated with, e.g., Monte Carlo sampling, which leads to noisy (approximate) objective and constraint evaluations. Instead of attempting to obtain high-quality estimates during each BO iteration, we propose a two-stage strategy to obtain tractable probabilistic guarantees. First, the constrained BO tuning procedure is repeated in a multi-start fashion (under multiple random seed points). Due to the random nature of the initialization and function evaluations, this generates a discrete set of ``good'' candidate tuning parameters that warrant further investigation. Then, we show how the optimal tuning parameter values from this set, along with a distribution-independent bound on performance and constraint violation probability, can be computed using non-convex scenario optimization \cite{campi2018general}. The advantages of the proposed automated controller tuning strategy 
are demonstrated on semibatch reactor case study 
controlled by economic nonlinear MPC. 

\textbf{Notation.} The set of non-negative and positive integers are denoted by $\mathbb{N}$ and $\mathbb{N}_+$, respectively. Given $a, b \in \mathbb{N}$ such that $a < b$, we let $\mathbb{N}_a^b = \{ a,a+1,\ldots,b \}$ denote the sequence of integers from $a$ to $b$. The $i^\text{th}$ element of a vector is denoted by $[x]_i$ and the $ij$ entry of a matrix $M$ is denoted by $[M]_{i,j}$. Given two column vectors $a$ and $b$, we let $(a,b) = [a^\top, b^\top]$.

\section{Problem Formulation}
\label{sec:problem-formulation}

We look to control the following discrete-time nonlinear system in the presence of uncertainty
\begin{align} \label{eq:sys}
x^+ = f(x,u,w),
\end{align}
where $x \in \mathbb{R}^{n_x}$ is the current state, $x^+$ is the successor state, $u \in \mathbb{R}^{n_u}$ is the control input, and $w \in \mathbb{R}^{n_w}$ is an unknown disturbance. We assume the state is perfectly measured and the disturbance takes values in a compact set $\mathcal{W} \subset \mathbb{R}^{n_w}$. The controlled system should satisfy general nonlinear state and input constraints of the form
\begin{align} \label{eq:constraints}
g_l(x, u, w) \leq 0,~~\forall l \in \mathbb{N}_1^{n_g},
\end{align}
where $g_l : \mathbb{R}^{n_x} \times \mathbb{R}^{n_u} \times \mathbb{R}^{n_w} \to \mathbb{R}$ are known functions and $n_g$ is the total number of constraints. We consider a generic control law $\kappa : \mathbb{R}^{n_x} \times \mathbb{R}^{n_\theta} \to \mathbb{R}^{n_u}$ that maps the state to the control input and is parametrized by $\theta \in \Theta \subseteq \mathbb{R}^{n_\theta}$, which represent the complete set of tuning parameters. Note that we do not impose any restrictions on the complexity of the control law, which be a non-convex function of $x$ or $\theta$. For example, $\kappa(x,\theta)$ could be an implicitly defined MPC law in which $\theta$ includes the prediction horizon, weights in the cost function, and/or constraint backoff terms.

Using the control law $u=\kappa(x,\theta)$ in the dynamics \eqref{eq:sys}, we obtain the closed-loop system
\begin{align} \label{eq:closed-loop-sys}
x(k+1) = f(x(k), \kappa(x(k), \theta), w(k)),
\end{align}
where $x(k)$ and $w(k)$ denote the state and disturbance at discrete time step $k$, respectively. A given trajectory of \eqref{eq:closed-loop-sys} is then defined by an admissible initial condition $x(0) \in \mathcal{X}_0$ and admissible disturbance sequence.
Instead of relying on asymptotic performance guarantees, the focus of this work is on finite-time closed-loop performance indicators that can be computed directly from closed-loop simulations. In particular, let $\delta = \{ x(0), w(0),\ldots,w(T-1) \}$ be the uncertain variables that define a closed-loop trajectory over a finite number of simulation time steps $T$, i.e.,
\begin{align}
z(\theta,\delta) = \{ & x(0),\kappa(x(0),\theta),w(0),\ldots,\\\notag
& x(T-1),\kappa(x(T-1),\theta),w(T-1),x(T) \}.
\end{align}
We assume $\delta \in \Delta$ where $\Delta$ is some probability space endowed with $\sigma$-algebra $\mathcal{D}$ and probability measure $\mathbb{P}$. The problem of interest in this work is to select the ``optimal'' controller design parameters via the optimization problem
\begin{subequations} \label{eq:hyper-opt}
\begin{align}
\label{eq:hyper-objective}
\theta^\star = \argmin_{\theta \in \Theta}&~~ \mathbb{E}_\delta \left[ F (\theta, \delta) \right], \\
\label{eq:hyper-constraints}
\text{s.t.} &~~ G(\theta,\delta) \leq 0, ~~ \forall \delta \in \Delta,
\end{align}
\end{subequations}
where $F :  \mathbb{R}^{n_\theta} \times \mathbb{R}^{n_\delta} \to \mathbb{R}$ is some specified performance indicator function, $n_\delta = n_x + Tn_w$, and $G :  \mathbb{R}^{n_\theta} \times \mathbb{R}^{n_\delta} \to \mathbb{R}$ is the worst-case constraint violation defined by
\begin{align}
G(\theta,\delta) = \max_{k \in \mathbb{N}_0^{T-1}}\max_{l \in \mathbb{N}_1^{n_g}}  g_l(x(k), \kappa(x(k),\theta), w(k)).
\end{align}
The objective function \eqref{eq:hyper-objective} is defined in terms of the expected value $\mathbb{E}_\delta\left[ F(\theta,\delta) \right] = \int_\Delta F(\theta,\delta) d\mathbb{P}$ and can be selected as any scalar function of the closed-loop trajectory $z(\theta, \delta)$. For example, we are often interested in some function of the final state $F(\theta, \delta) = \phi(x(T))$ or an average cost
\begin{align}
F(\theta,\delta) = \frac{1}{T} \sum_{k=0}^{T-1} \ell(x(k), \kappa(x(k), \theta), w(k)),
\end{align}
where $\ell : \mathbb{R}^{n_x} \times \mathbb{R}^{n_u} \times \mathbb{R}^{n_w} \to \mathbb{R}$. The constraints \eqref{eq:hyper-constraints} are enforced for all possible uncertainty values $\delta \in \Delta$ in which the uncertainty set $\Delta$ is, in most situations, a continuous set containing an infinite number of instances -- resulting in \eqref{eq:hyper-opt} being a \textit{semi-infinite} optimization problem that is difficult to solve, especially when $F(\cdot,\delta)$ or $G(\cdot,\delta)$ are non-convex for any $\delta \in \Delta$. Note that no additional assumptions are imposed on $\Theta$, which can have a mixture of continuous, discrete, and categorical components.

Since no closed-form solution exists for the controller tuning optimization problem \eqref{eq:hyper-opt}, we look to take advantage of the so-called \textit{scenario approach} \cite{campi2009scenario} that derives a probabilistic relaxation by replacing the expected value and worst-case operators with random sample-based approximations. The details of non-convex scenario theory and how it can be applied to \eqref{eq:hyper-opt} are discussed next.

\section{Scenario-based relaxation of controller tuning optimization problem}
\label{sec:solution-method}

\subsection{Non-convex scenario optimization}

Let $(\Delta^N, \mathcal{D}^N, \mathbb{P}^N)$ be the $N$-fold Cartesian product of $\Delta$ equipped with the product $\sigma$-algebra $\mathcal{D}^N$ and the product probability 
$\mathbb{P}^N = \mathbb{P} \times \cdots \times \mathbb{P}$. Thus, a point in $(\Delta^N, \mathcal{D}^N, \mathbb{P}^N)$ is a sample $(\delta^1,\ldots,\delta^N)$ of $N$ components extracted independently from $\Delta$ according to the same probability $\mathbb{P}$. Each $\delta^i$ is referred to as a ``scenario'' and represents the collection of uncertain external variables that define the closed-loop system. For any sample $(\delta^1,\ldots,\delta^N)$, we can construct the following scenario approximation to \eqref{eq:hyper-opt}
\begin{subequations} \label{eq:scenario-approximation-hyper}
\begin{align}
\min_{\theta \in \Theta, \xi \geq 0} &~~ \rho \| \xi \| + \frac{1}{N}\sum_{i=1}^N F (\theta, \delta^i), \\
\text{s.t.} &~~ G(\theta,\delta^i) \leq \xi, ~ \forall i = 1,\ldots,N,
\end{align}
\end{subequations}
where $\xi \in \mathbb{R}$ is a slack variable that represents the maximum constraint violation over all possible uncertainty values and $\rho > 0$ is the penalty weight. We soften constraints \eqref{eq:hyper-constraints} to guarantee the existence of a feasible solution, which is an important assumption in the theory of scenario optimization (discussed in more detail below). Even though \eqref{eq:scenario-approximation-hyper} allows constraints to possibly be violated, this violation results in a regret $\xi$ that is added to the original cost function. The parameter $\rho$ is used to achieve a reasonable tradeoff between minimizing the original cost and the ``regret'' for constraint violation. For large enough $\rho$ values, \eqref{eq:scenario-approximation-hyper} is an \textit{exact penalty function} that results in the same solution \eqref{eq:hyper-opt} when the original problem is feasible. In particular, $\rho$ must be larger than the dual norm of the optimal Lagrange multiplier for the constraint $\max_{i \in \mathbb{N}_1^N}G(\theta,\delta^i)$ (see \cite[Theorem 1]{kerrigan2000soft}). Since this condition is difficult to verify \textit{a priori}, we must often resort to a heuristic procedure for selecting $\rho$ in practice. 

Let $y = (\theta, \xi) \in \mathcal{Y} = \Theta \times \mathbb{R}_{\geq 0}$ denote the decision variables in \eqref{eq:scenario-approximation-hyper} and define the associated constraints
\begin{align} \label{eq:scenario-sets}
\mathcal{Y}_\delta = \{ y \in \mathcal{Y} : G(\theta,\delta) - \xi \leq 0 \},
\end{align}
for every $\delta \in \Delta$. Since \eqref{eq:scenario-approximation-hyper} is only enforcing constraints at a fixed number of scenarios, it is not possible to establish a 100\% guarantee of constraint satisfaction. Instead, we can establish a bound on the probability that constraints will be violated, which is defined as follows.

\begin{definition}
The \textit{violation probability} of a given decision variable $y \in \mathcal{Y}$ is defined as
\begin{align}
\mathbb{V}(y) = \mathbb{P} \{ \delta \in \Delta : y \not\in \mathcal{Y}_\delta \}.
\end{align}
For a reliability parameter $\varepsilon \in (0,1)$, $y \in \mathcal{Y}$ is said to be $\varepsilon$-feasible if $\mathbb{V}(y) \leq \varepsilon$. $\hfill \triangleleft$
\end{definition}

Ideally, we could exactly compute the violation probability at the solution to \eqref{eq:scenario-approximation-hyper}; however, two problems remain: (i) \eqref{eq:scenario-approximation-hyper} is a non-convex problem such that finding its globally optimal solution is a difficult task and (ii) $\mathbb{V}(y)$ is defined in terms of an infinite-dimensional integral over the probability space $\Delta$ for any $y$. Both of these challenges are addressed by the theory established in \cite{campi2018general} for non-convex scenario decision problems, as briefly recalled below.

Assume an algorithm $\mathcal{A}_N : \Delta^N \to \mathcal{Y}$ exists that maps the samples $(\delta^1,\ldots,\delta^N)$ to a solution $y_N^\star = \mathcal{A}_N(\delta^1,\ldots,\delta^N)$. Due to its dependence on the randomly drawn scenarios, $y_N^\star$ is random, implying $\mathbb{V}(y_N^\star)$ is a random variable defined over $\Delta^N$. The main idea behind scenario optimization is then to establish confidence bounds for the $\varepsilon$-feasibility of $y_N^\star$ by analyzing the distribution of $\mathbb{V}(y_N^\star)$. This problem has been heavily studied in the convex case and relies on \textit{support constraints}, which are defined as follows. 

\begin{definition} \label{def:subsample}
Given $( \delta^1,\ldots,\delta^N ) \in \Delta^N$, a \textit{support subsample} $S$ for $( \delta^1,\ldots,\delta^N )$ is a $k$-tuple of elements ($k \in \mathbb{N}_1^N$) extracted from $( \delta^1,\ldots,\delta^N )$, i.e., $S = (\delta^{i_1},\ldots,\delta^{i_k})$ with $i_1 < \cdots < i_k$ that yields the same solution as the full sample
\begin{align}
\mathcal{A}_k(\delta^{i_1},\ldots,\delta^{i_k}) = \mathcal{A}_N(\delta^1,\ldots,\delta^N).
\end{align}
A support subsample $S$ is said to be \textit{irreducible} if no element can be further removed from $S= (\delta^{i_1},\ldots,\delta^{i_k})$ without changing the solution. $\hfill \triangleleft$
\end{definition}

Note that support constraints are closely related to active constraints -- in fact support constraints are always active constraints, but the converse is not necessarily true as shown in \cite[Fig. 2]{campi2018wait}. Since the number of support constraints is at most $n_y$ for an $n_y$-dimensional convex optimization problem, this has been commonly used to establish \textit{a priori} bounds on $\mathbb{V}(y_N^\star)$. As this result no longer holds in non-convex optimization, we must rely on an \textit{a posteriori} determination that, when combined with the following theorem, can give us the desired confidence bound on the violation probability. 

\begin{theorem}[\cite{campi2018general}] \label{thm:scenario}
Suppose there exists a solver $\mathcal{A}_N$ that provides a (possibly suboptimal) unique solution to \eqref{eq:scenario-approximation-hyper}. Let $\beta \in (0,1)$ be a desired confidence parameter for $\mathbb{V}(y_N^\star)$ and let $\varepsilon : \{ 0,\ldots,N \} \to [0,1]$ be a function satisfying
\begin{align} \label{eq:sum-varepsilon}
\sum_{k=1}^N \begin{pmatrix}
N \\
k
\end{pmatrix} (1 - \varepsilon(k))^{N - k} = \beta,~~\varepsilon(N)=1.
\end{align}
Suppose that an algorithm $\mathcal{B}_N : \Delta^N \to 2^{\{ 1,\ldots,N \}}$ is available that can select a support subsample for \eqref{eq:scenario-approximation-hyper} and, in addition, let $s^\star_N = | \mathcal{B}_N(\delta^1,\ldots,\delta^N) |$ be the size/length of the support subsample. Then, it holds that
\begin{align} \label{eq:scenario-result}
\mathbb{P}^N \{ \mathbb{V}(y_N^\star) > \varepsilon(s_N^\star) \} \leq \beta,
\end{align}
 for any $\mathcal{A}_N$, $\mathcal{B}_N$, and probability $\mathbb{P}$. $\hfill \blacksquare$
\end{theorem}

Let us highlight some important points in Theorem \ref{thm:scenario}. First, the result is quite general as it holds for any uncertainty distribution and any solution method $\mathcal{A}_N$ (as long as it provides a unique result). 
Second, there are two levels of probability in \eqref{eq:scenario-result} -- since probabilities must sum to 1, 
we can rearrange this to $\mathbb{P}^N \{ \mathbb{V}(y_N^\star) \leq \varepsilon(s_N^\star) \} > 1 - \beta$. Thus, the inner level is the claim that $y_N^\star$ is $\varepsilon(s_N^\star)$-feasible and the outer level states this claim holds true with probability at least $1 - \beta$. The value of $\beta$ is chosen by the user and often selected to be practically zero with a common default value of $10^{-6}$. Given $\beta$, the resulting $\varepsilon(\cdot)$ must be chosen to satisfy \eqref{eq:sum-varepsilon}. The values $\varepsilon(k)$ distribute $\beta$ across different possible observations of $k \in \mathbb{N}_1^N$ support constraints. Whenever $\beta$ is split equally among the $N$ terms, we can derive
\begin{align} \label{eq:epsilon-explicit}
\varepsilon(k) = \begin{cases}
1 &\text{if~} k=N, \\
1 - \sqrt[N-k]{\frac{\beta}{N {N\choose k}}} &\text{otherwise}.
\end{cases}
\end{align}
The weak (logarithmic) dependence on $\beta$ is an important advantage of this bound, though many other bounds satisfying \eqref{eq:sum-varepsilon} are possible as discussed in detail in \cite{campi2018general}. 

\subsection{Practical application to discrete optimization problems}
\label{subsec:practical-discrete-solution}

To apply Theorem \ref{thm:scenario}, we need to construct an algorithm $\mathcal{B}_N$ that can identify support subsamples (Definition \ref{def:subsample}) for the scenario optimization \eqref{eq:scenario-approximation-hyper}. A trivial choice is an algorithm that always returns $N$ (as the full sample is obviously a support subsample as well), however, this results in a useless bound of $\mathbb{P}^N \{ \mathbb{V}(y_N^\star) > 1 \} \leq \beta$. 
Therefore, it is important to find smaller support subsamples to get meaningful results; the least conservative result being the irreducible support subsample of minimal length -- often referred to as the \textit{essential set}. Identification of the essential set requires (in the worst-case) enumerating all $2^N$ possible permutations of the scenario constraints and solving the resulting $2^N$ non-convex problems, which quickly becomes computationally intractable \cite{geng2019general}. To avoid this exponential growth, we can use the following greedy algorithm that requires solving only $N$ instead of $2^N$ non-convex problems \cite{campi2018general}:
\begin{enumerate}[1)]
\item Set $L_s \leftarrow (\delta^1,\ldots,\delta^N)$ and compute $y_N^\star \leftarrow \mathcal{A}_N(L_s)$.
\item For all $i = 1,\ldots,N$
\begin{itemize}
\item Set $L_s' \leftarrow L_s \setminus \delta^i$ and compute $\bar{y} \leftarrow \mathcal{A}_{| L_s' |}(L_s')$.
\item If $\bar{y} = y_N^\star$, then set $L_s \leftarrow L_s'$. 
\end{itemize}
\item Output the set $\{ i_1,\ldots i_k \}$, $i_1 < \ldots < i_k$, of the indexes of the elements in $L_s$. 
\end{enumerate}
This choice for $\mathcal{B}_N$ is guaranteed to find an irreducible support subsample. Furthermore, under an additional non-degeneracy assumption on the scenario program, this algorithm will return the (unique) essential set as recently shown in \cite[Theorem 4]{geng2019general}. However, this still requires solving many variations of \eqref{eq:scenario-approximation-hyper} defined in terms of expensive-to-evaluate closed-loop simulations.
Assuming a constant number of $N_\text{max}$ iterations, we would have to perform a worst-case total of $N_\text{max} N^2$ closed-loop simulations can be prohibitive for a practical controller tuning strategy. 

An important case where we can simplify this procedure is when the set  $\Theta = \{ \theta^1,\ldots,\theta^{N_\theta} \}$ is composed of (or approximated by) a collection of $N_\theta$ discrete values. In this case, we can directly evaluate the cost and constraint functions and store them in matrices $\mathbf{F}$ and $\mathbf{G}$
\begin{align} \label{eq:matrix-scenario-cost-constraint}
[\mathbf{F}]_{ij} = F(\theta^j, \delta^i), ~~[\mathbf{G}]_{ij} = G(\theta^j, \delta^i),
\end{align}
for all $i \in \mathbb{N}_1^N$ and $j \in \mathbb{N}_1^{N_\theta}$. Since these can be evaluated simultaneously, we only need a total of $N_\theta N$ evaluations, which is linear in the number of scenarios. By computing and storing these values \textit{a priori}, we can significantly reduce the cost of $\mathcal{B}_N$. In particular, the scenario optimization \eqref{eq:scenario-approximation-hyper} reduces to the following discrete optimization problem
\begin{align} \label{eq:scenario-finite-opt}
\min_{j \in \mathbb{N}_1^{N_\theta}} \left\lbrace \rho \max_{i \in \mathbb{N}_1^N} ([\mathbf{G}]_{ij},0) + \frac{1}{N} \sum_{i=1}^N [\mathbf{F}]_{ij} \right\rbrace.
\end{align}
The major cost is then populating the matrices \eqref{eq:matrix-scenario-cost-constraint} while $\mathcal{A}_N$ can straightforwardly be selected to globally optimize \eqref{eq:scenario-finite-opt} using standard operations on matrices. However, it is important to note that we are limiting ourselves to consider only a finite number of parameters. Thus, even if $\Theta$ is composed only of discrete variables (e.g., prediction and control horizon in MPC), the number of candidates will grow exponentially with dimension as we must consider all possible combinations of these variables. Thus, it is important to develop an effective procedure for filtering values out of $\Theta$ such that only the most viable tuning parameters are considered in the formal optimization procedure. We discuss an effective selection strategy next based on recent advancements in constrained Bayesian optimization.

\section{Generation of Candidate Tuning Parameters using Constrained Bayesian Optimization}
\label{sec:candidate-cbo}

We can write problem \eqref{eq:hyper-opt} in the following equivalent form
\begin{align} \label{eq:black-box-opt}
\min_{\theta \in \Theta} ~ \mathcal{J}(\theta) ~~ \text{s.t.}~~ \mathcal{C}(\theta) \geq 0,
\end{align}
where $\mathcal{J}(\theta) = \mathbb{E}_\delta \left[ F (\theta, \delta) \right]$ and $\mathcal{C}(\theta) = \mathbb{P}_\delta[ G(\theta,\delta) \leq 0 ] - 1$ is the probabilistic statement of the robust constraint \eqref{eq:hyper-constraints}. We cannot exactly evaluate the functions $\mathcal{J}(\cdot)$ and $\mathcal{C}(\cdot)$ for even a single $\theta \in \Theta$, as uncertainty must be propagated through the generally nonlinear closed-loop dynamics \cite{paulson2018nonlinear}. We can, however, apply Monte Carlo sampling (MCS) to approximate the closed-loop performance indicators as follows
\begin{subequations} \label{eq:saa-estimates}
\begin{align}
\mathcal{J}(\theta) &\approx \hat{\mathcal{J}}(\theta) = M^{-1} \textstyle \sum_{i=1}^M F(\theta, \tilde{\delta}^i), \\
\label{eq:saa-constraints}
\mathcal{C}(\theta) &\approx \hat{\mathcal{C}}(\theta) = M^{-1} \textstyle \sum_{i=1}^M \mathbf{1}_{(-\infty,0]}( G(\theta,\tilde{\delta}^i )) - 1,
\end{align}
\end{subequations}
where $M$ is the number of samples, $\mathbf{1}_A(x)$ is the indicator function over the set $A$, and $(\tilde{\delta}^1,\ldots, \tilde{\delta}^M)$ are i.i.d. samples of the uncertainty and the $\tilde{\cdot}$ is used to differentiate these samples from those used in the scenario optimization \eqref{eq:scenario-approximation-hyper}. To derive \eqref{eq:saa-constraints}, we first substituted the expression $\mathbb{P}_\delta[ G(\theta,\delta) \leq 0 ] = \mathbb{E}_\delta[ \mathbf{1}_{(-\infty,0]}(G(\theta,\delta)) ]$. This reformulation is important because the MCS estimates in \eqref{eq:saa-estimates} are unbiased for any $M \in \mathbb{N}_+$, which is an implicit assumption in most simulation optimization methods that should be satisfied. 

Problem \eqref{eq:black-box-opt} is inherently challenging due to its black-box nature and the presence of noisy and expensive function evaluations. Thus, we look to apply an algorithm that can more systematically explore the tuning parameter space $\Theta$ relative to random or grid-based search methods. Bayesian optimization (BO) is a particularly powerful family of algorithms for varied black-box design problems that has been successfully applied in several application domains \cite{shahriari2015taking,snoek2012practical,marchant2012bayesian}
The main idea behind BO is to build a surrogate model for the objective function using a set of $n$ observations denoted by $\mathcal{D}_n = \{ (\theta_{1:n}, \hat{\mathcal{J}}_{1:n}) \}$ where $\hat{\mathcal{J}}_{1:n} = (\hat{\mathcal{J}}(\theta_1),\ldots,\hat{\mathcal{J}}(\theta_n))$. This surrogate model is used to compute a posterior distribution for the objective -- this posterior is combined with an \textit{acquisition function} (AF) to determine which next tuning parameter value should be sampled. The AF is chosen in such a way that uncertainty in the predictions of $\mathcal{J}(\theta)$ can be traded off with performance (in terms of the best mean value) at unexplored points $\theta \in \Theta \setminus \theta_{1:n}$. The objective realization at the next sample $\theta_{n+1}$ is added to the data $\mathcal{D}_{n+1} = \mathcal{D}_n \cup \{ (\theta_{n+1}, \hat{\mathcal{J}}_{n+1}) \}$ that can be used to update the surrogate model. This entire process is repeated until convergence or (more commonly) the maximum number of iterations have been executed. Global convergence of BO has been analyzed \cite{bull2011convergence} and there are various open-source software implementations available, e.g., \cite{martinez2014bayesopt}. 

A wide-variety of probabilistic surrogate models have been used to represent the objective function in BO including parametric and non-parametric model types. The latter is often preferred due to their ability to represent any function given a sufficiently large data set. Gaussian process (GP) models are the most commonly used since their posterior distribution can be derived analytically \cite{jones1998efficient}. There are several choices for the AF, denoted by $\alpha_n(\theta ; \mathcal{D}_n)$ at iteration $n$, with the default selection usually the expected improvement (EI) criteria \cite{shahriari2015taking}.
Regardless of the choice of AF, it does not account for constraints. Thus, an intuitive extension is to define improvement as occurring only when constraints are satisfied. This implies that our next sampling point $\theta_{n+1}$ can be obtained by solving the following optimization problem
\begin{align} \label{eq:acquisition-function}
\theta_{n+1} = \text{argmax}_{\theta \in \Theta} \left\lbrace \alpha_n(\theta ; \mathcal{D}_n) \mathbb{P} [ \mathcal{C}(\theta) \geq 0 \mid \mathcal{D}^c_n ] \right\rbrace,
\end{align}
where $\mathcal{D}^c_n = \{ (\theta_{1:n}, \hat{\mathcal{C}}_{1:n}) \}$ are the noisy constraint evaluations. The probability term can again be analytically computed for GP models \cite{hernandez2016general}. Since the terms in \eqref{eq:acquisition-function} are cheap to evaluate using the GP models, this maximization can be carried out efficiently. When we do not have any feasible data points, it can be useful to neglect the AF factor and instead maximize the probability of constraint satisfaction -- this search is purely exploitative and will discover a particular region of $\Theta$ is feasible or its probability will drop and the algorithm will move onto a more promising region. 

In addition to selecting the AF, one must choose the number of seed points $N_\text{seed}$ and maximum number of iterations $N_\text{max}$. Even though these values can have a significant affect on solution quality, this is mitigated in this work by running the complete BO scheme $N_\theta$ times with different randomly selected seed points
This set of returned solutions represents the collection of candidate controller tunings $\tilde{\Theta} = \{ \hat{\theta}_{N_{\text{max}},i} \}_{i=1}^{N_\theta}$, where $\hat{\theta}_{N_{\text{max}},i}$ denotes the best solution found during the $i$th BO run, that replaces the set $\Theta$ in the scenario optimization problem \eqref{eq:scenario-approximation-hyper}. 


\section{Numerical Illustration}
\label{sec:numerical-results}

The effectiveness of the proposed controller tuning method is demonstrated on a jacketed semibatch reactor with exothermic reaction $A + B \to C$ \cite{paulson2019efficient}.
The dynamics are described by a set of nonlinear ordinary differential equations
\begin{subequations} \label{eq:semibatch-ode}
\begin{align}
\dot{V} &= \dot{V}_\text{in}, \\
\dot{c}_A &= -F_\text{in} c_A - kc_A c_B, \\
\dot{c}_B &= F_\text{in} (c_{B,\text{in}} - c_B) - kc_A c_B + [w]_1, \\
\dot{c}_C &= -F_\text{in} c_C + kc_A c_B, \\
\dot{T}_r &= F_\text{in}(T_\text{in} -T_r) - \frac{\alpha (T_r - T_J)}{\rho V c_p} - \frac{k c_A c_B H}{\rho c_p}, \\
\dot{T}_J &= F_{J,\text{in}} (T_{J,\text{in}} - T_J) + \frac{\alpha (T_r - T_J)}{\rho V_J c_p} + [w]_2, \\
\dot{T}_{J,\text{in}} &= \tau_c^{-1} (T_{J,\text{in},\text{set}} - T_{J,\text{in}}) + [w]_3,
\end{align}
\end{subequations}
where $V$ is the reactor volume; $c_i$ is the concentration of species $i \in \{ A, B, C \}$; $T_r$ is the reactor temperature; $T_J$ is the jacket temperature; $T_{J,\text{in}}$ is the jacket inlet temperature; $\dot{V}_\text{in}$ is the feed rate of component $B$; and $T_{J,\text{in},\text{set}}$ is the setpoint of the jacket inlet temperature. The heat transfer area is denoted by $A = 2V/r + \pi r^2$, where $r$ is the reactor radius and $F_\text{in} = \dot{V}_\text{in}/V$. The control inputs are $u = (\dot{V}_\text{in}, T_{J,\text{in},\text{set}})$, the state vector is $x = (V, c_A, c_B, c_C, T_r, T_J, T_{J,\text{in}})$, and the disturbances $w \in \mathbb{R}^3$ are additive in the right-hand side of the $c_B$, $T_J$, and $T_{J,\text{in}}$ equations. The random disturbances are assumed to be uniformly distributed within $w \in [-0.5,0.5] \times [-0.05,0.05]^2$. The model parameters, initial conditions, and input constraints are given in \cite[Table 2]{paulson2019efficient}. 
There are two safety constraints of the form \eqref{eq:constraints} on the reactor temperature
\begin{align}
g_1(x,u,w) = [x]_5 - 326, ~ g_2(x,u,w) = 322 - [x]_5.
\end{align}
The control objective is to maximize moles of $C$ at the end of the batch $t_f = 1200$ s, i.e., $F(\theta,\delta) = -c_C(t_f) V(t_f)$.
Control inputs are updated every $\delta t = 30$ s such that $T = 40$. We design a nonlinear MPC controller that solves an optimization at every time $t_k$ for discrete times $k \in \mathbb{N}_0^{T-1}$
\begin{align}
\min_{x_{i|k}, u_{i|k}} &~~ \textstyle\sum_{i=0}^{\theta_p - 1} (-[x_{i|k}]_1 [x_{i|k}]_3) + V_f(x_{\theta_p|k}), & \\\notag
\text{s.t.} &~~ x_{i+1|k} = f_{\theta_d} (x_{i|k}, u_{i|k}, \hat{w}_{i|k}), & i \in \mathbb{N}_0^{\theta_p-1}, \\\notag
&~~ g(x_{i|k}, u_{i|k}, w_{i|k}) \leq -\theta_b, & i \in \mathbb{N}_0^{\theta_p-1}, \\\notag
&~~ u_{i|k} \in \mathbb{U}, & i \in \mathbb{N}_0^{\theta_p-1}, \\\notag
&~~ x_{0|k} = x(k), \\\notag
&~~ x_{\theta_p | k} \in \mathbb{X}_f,
\end{align}
where $x_{i|k}$, $u_{i|k}$, and $\hat{w}_{i|k} = 0$ are the predicted state, input, and disturbance values at $i$ steps ahead of current discrete time $k$; $x(k)$ is the measured state at time $t_k$; $V_f(x) = 0$ and $\mathbb{X}_f = \mathbb{R}^n$  are the terminal cost and constraints (neglected in this work but can be selected to ensure nominal stability as discussed in \cite{rawlings2009model}); $\mathbb{U} \subset \mathbb{R}^2$ are the control input constraints; and $\theta = (\theta_b, \theta_p, \theta_d)$ are the tunable parameters. Here, parameters $\theta_b \in [0, 0.5]^2 \subset \mathbb{R}^2$ are constraint backoff values that can be selected to improve the inherent robustness guarantees in MPC, as shown in, e.g., \cite{paulson2018shaping,paulson2018nonlinear}, $\theta_p \in \mathbb{N}_5^{20}$ is the integer-valued prediction horizon, and $\theta_d \in \{ \small\texttt{ForwardEuler}, \texttt{RK4}, \texttt{Collocation}, \texttt{ImplicitEuler} \}$ is a categorical variable representing the type of discretization scheme used to approximate the differential equations \eqref{eq:semibatch-ode}. The value $\small\texttt{RK4}$ represents a 4th order Runge Kutta scheme while $\small\texttt{Collocation}$ represents orthogonal collocation on finite elements, as discussed in \cite{biegler2007overview}. 
All MPC problems are solved with CasADi \cite{andersson2012casadi} and IPOPT \cite{wachter2006implementation}. The complete code used for this letter is available for download at \url{https://github.com/joelpaulson/LCSS_DataDrivenScenarioOptimization}.

\begin{figure}[t]
\centering
\includegraphics[width=0.8\linewidth]{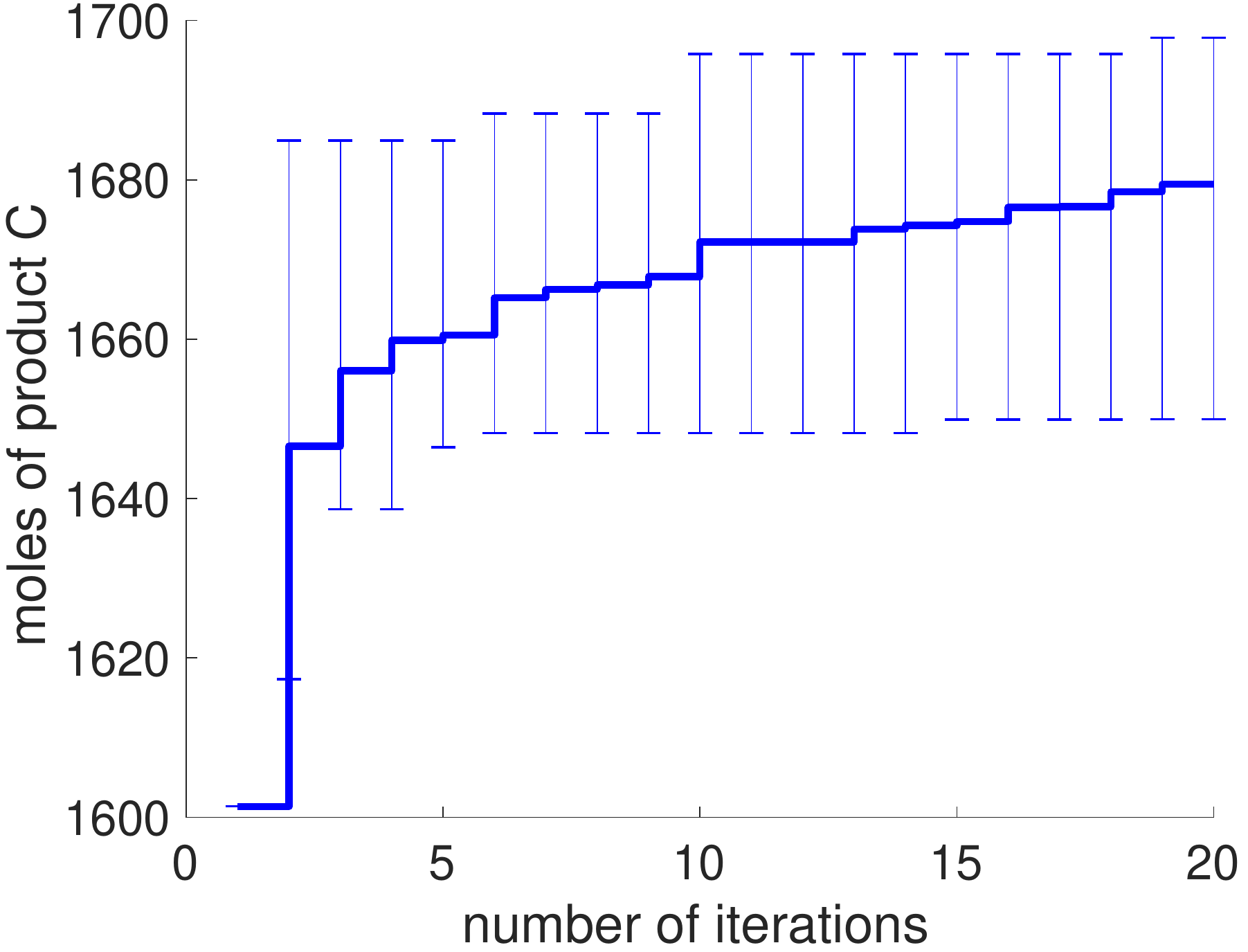}
\caption{Maximum moles of C versus iteration number for the EI acquisition function. The BO algorithm was repeated 15 times. The mean is shown in bold and the min/max observed values are shown with errorbars.}
\label{fig:moles-product-BO}
\end{figure}

First, we generate a set of candidate tuning parameters $\tilde{\Theta}$ using the constrained BO algorithm presented in Section \ref{sec:candidate-cbo}, which can be implemented using the $\mathtt{bayesopt}$ function in MATLAB. To ensure a reasonable computational cost, we set $N_\text{seed} = 5$, $N_\text{max} = 20$, $M=3$, and the EI AF, which takes $\sim 2.5$ min to return a solution on a MacBook Pro with 32 GB of RAM and 2.3 GHz Intel i9 processor. One feasible seed point $\theta = (0.5, 0.5, 5, \small\texttt{ForwardEuler})$ was provided with maximum backoffs, which results in overly conservative performance. This process was repeated $N_\theta = 15$ times to populate $\tilde{\Theta}$, with the best, worst, and average performance across all BO iterations shown in Fig. \ref{fig:moles-product-BO}. Notice how performance consistently improves as the number of iterations increases, however, there is still a significant amount of variability in the solutions. 

Since it is not obvious which tuning parameter in $\tilde{\Theta}$ is best, we apply the scenario approach for discrete sets shown in \eqref{eq:scenario-finite-opt}. For $N=750$, we found the solution to be $\theta^\star_{750} = (0.495, 0.031, 7, \small\texttt{RK4})$ and $\xi^\star_{750} = 0$ using a large constraint violation penalty $\rho = 10^6$. We can evaluate the $\epsilon$-feasibility of this solution using Theorem \ref{thm:scenario}. First, we apply the greedy algorithm for $\mathcal{B}_N$ in Section \ref{subsec:practical-discrete-solution} to find a support subsample of length $s^\star_{750} = 1$. Then, we select $\beta = 10^{-6}$ and use in \eqref{eq:epsilon-explicit} to establish $\varepsilon(s^\star_{750}) = \varepsilon(1) = 0.0355$. This implies that $\mathbb{P}[ G(\theta^\star_{750}, \delta) > 0 ] \leq 3.55\%$, i.e., the closed-loop system does not violate the safety constraints with probability at least $96.45\%$ with confidence $1 - \beta \approx 1$. 

\begin{figure}[t]
\centering
\includegraphics[width=0.8\linewidth]{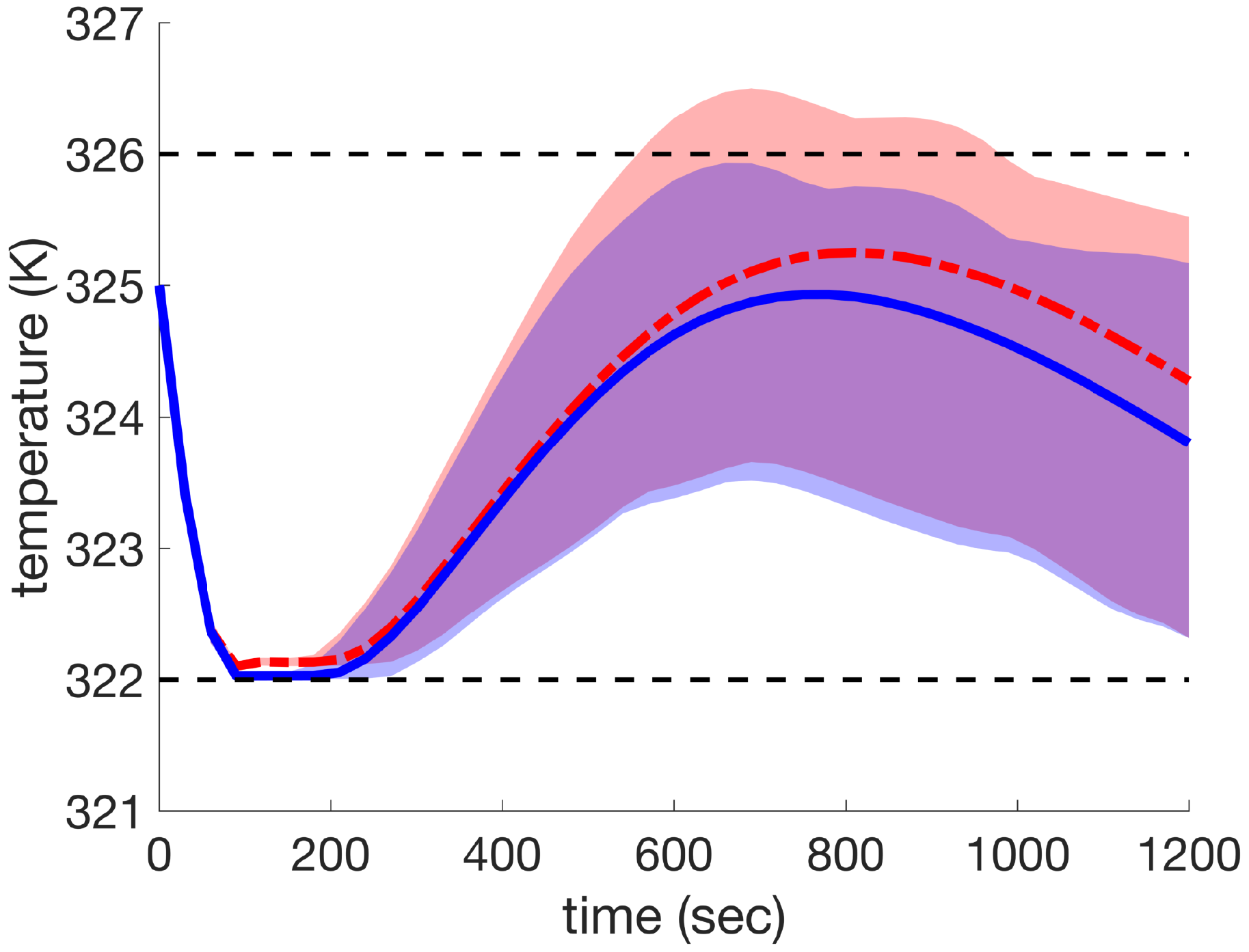}
\caption{Closed-loop temperature profiles using tuning parameters from $\tilde{\Theta}$ that optimize the scenario program \eqref{eq:scenario-finite-opt} (blue) and result in the largest constraint violation (red). The shaded cloud regions cover the minimum and maximum values for the 750 random scenarios, while the solid blue and dashed dotted red lines represent the average values at each time.}
\label{fig:temperature-scenario}
\end{figure}

The closed-loop temperature profiles for the 750 scenarios generated by $\kappa(x, \theta^\star_{750})$ are shown in Fig. \ref{fig:temperature-scenario}. We observe $\kappa(x, \theta^\star_{750})$ forces the system near the minimum and maximum temperature bounds (without violating them). Note that the lower bound backoff is considerably smaller than the upper bound backoff. This type of result would not have been easy to determine by trial-and-error. For comparison purposes, we also plotted results for $\kappa(x, \theta_{wc})$, where $\theta_{wc} \in \tilde{\Theta}$ is the BO-identified tuning parameter that provided the worst-case constraint violation. Our results suggest one cannot simply trust BO -- additional validation or optimization methods are needed to protect against outlier results. 

\section{Conclusions}
\label{sec:conclusions}

This paper presents a novel method for providing probabilistic closed-loop performance guarantees in automatic optimization-based controller tuning for generic control structures. The proposed approach applies non-convex scenario optimization theory to evaluate a distribution-free bound on expected performance and the probability of constraint violation. To reduce computational cost, this theory is applied to a discrete set of candidate tuning parameters obtained from repeated runs of a constrained Bayesian optimization (CBO) algorithm. Future work will look to improve the convergence rate of CBO using more sophisticated uncertainty propagation techniques that can reduce variance in the estimates of the closed-loop objective and constraints. 

\bibliographystyle{ieeetr}
\bibliography{references}

\end{document}